# Interface Probing by Dielectric Frequency Dispersion in Carbon Nanocomposites

*Yuhan Li, Faxiang Qin\*, Huan Wang, Hua-Xin Peng*

Y.H. Li, Prof. F.X. Qin, Dr. H. Wang, Prof. H.X. Peng
Institute for Composites Science Innovation (InCSI), School of Material Science and Engineering, Zhejiang University, Zheda Road, Hangzhou, 310027, China
E-mail: faxiangqin@zju.edu.cn

Interfaces remain one of the major issues in limiting the understanding and designing polymer nanocomposites due to their complexity and pivotal role in determining the ultimate composites properties. In this study, we take multi-walled carbon nanotubes/silicone rubber nanocomposites as a representative example, and have for the first time studied the correlation between high-frequency dielectric dispersion and static/dynamic interfacial characteristics. We have found that the interface together with other meso-structural parameters (volume fraction, dispersion, agglomeration) play decisive role in formulating the dielectric patterns. The calculation of the relaxation times affords the relative importance of interfacial polarization to dipolar polarization in resultant dielectric relaxation. Dielectric measurements coupled with cyclic loading further reveals the remarkable capability of dielectric frequency dispersion in capturing the evolution of interfacial properties, such as a particular interface reconstruction process occurred to the surfactant-modified samples. All these results demonstrate that high-frequency dielectric spectroscopy is instrumental to probing both static and dynamic meso-structural characteristics, especially effective for the composites with relative weak interfaces which remains a mission impossible for many other techniques. The insights provided here based on the analyses of dielectric frequency dispersion will pave the way for optimized design and precise engineering of meso-structure in polymer nanocomposites.

# 1. Introduction

Polymer-based nanocomposites have been extensively investigated as multifunctional and high performance materials and their applications have been widely extended due to their tailored properties including damping capacity, thermal conductivity, flame retardancy, electrical conductivity and so forth.[1-4] With the addition of nano-fillers, structural characteristics and fundamental physics involved can be distinguished from traditional polymer composites.[5] One significant difference is the predominant influence of interfaces in nanocomposites on the overall material properties. As interfaces become more spatially extensive and complex with reducing filler size,[6] characterizing and designing interfaces in nanocomposites have become a requisite for further optimizing materials performance. Much research efforts have been devoted to digging into the nature of interfaces in nanocomposites. Following this idea, microscopic techniques have been applied for direct observation of interface. For example, transmission electron microscopy (TEM) had been used for analyzing the microstructure of interface.[7] Forster resonance energy transfer and (FRET) and Laser scanning confocal microscopy (LSCM) were integrated to characterize the formation of interface at nanoscale.[8] It is believed that interfacial properties could influence mechanical properties greatly considering their essential role in load transfer from matrix to fillers. Increase of fracture toughness, Young's modulus and tensile strength have been realized in covalently bonded graphene platelets (GnPs)/epoxy nanocomposites comparing with the unmodified samples.[7] Molecular simulation results have shown that a low density of functionalization at the interface area influenced the shear strength of nanocomposites remarkably.[9] As researchers strive to achieve the full potential of nanocomposites, exploring the coupling mechanism between functionality and interfacial properties becomes a necessity. Peng et al. used electrostatic force microscopy (EFM) for characterization of interface and found the decrease of local permittivity contributed by interface in nanocomposites.[10] Lewis et al. explained that as length scale decreased to 200 nm and below, nanodielectrics and interface became inextricable and interfacial properties strongly influenced overall dielectric properties.[11] Nan et al. formulated a simplified equation for predicting the thermal conductivity in carbon nanotube composites, where it had been shown that thermal conductivity of interface influences the effective thermal properties dramatically.[12] However, key issues still exist in the research for interfaces in nanocomposites: (i) it remains a challenge to characterize relatively weak interfaces; (ii) previous research mainly emphasized on the interface itself and little attention was given to investigating the interplay between interfaces and other structural characteristics; (iii) the intrinsic effects of specific interface conditions in determining the macroscopic functionality of materials remain obscure, making it difficult to apply the principles obtained from individual study to the design and fabrication of general nanocomposites. It appears that the existing single method cannot meet the need for sufficiently comprehending and utilizing interfaces; as such, diversified research methods are desirable to be explored, particularly those that can approach rich information of interfacial characteristics from somewhat unique or holistic perspective.

Apart from direct observation of interface, spectroscopic methods (e.g., Raman spectroscopy, electron energy-loss spectroscopy)[13-15] have also been adopted to analyze interfacial properties, where external physical fields are applied to the material to detect the corresponding change of structural details linked to interface. Dielectric spectroscopy (DS) has long been used for measuring dielectric parameters and analyzing dielectric response of bulk materials, with its extended testing frequency range nowadays and its sensitivity to intermolecular interactions and dynamic process.[16] Herein, we propose capitalizing on the strength of DS in characterizing structural properties at multiple length scales corresponding to its wide frequency range for studying polymer nanocomposites. In particular, we attempt to demonstrate that dielectric spectroscopy is a promising tooling for characterizing interfacial properties and other structural characteristics of nanocomposites to reveal the real effects of interfaces in functional nanocomposites.

In this paper, we choose multi-walled carbon nanotubes (MWCNTs) as functional fillers and silicone elastomer as matrix considering the convenience of tuning interfacial properties based on previous studies on covalent or non-covalent modification methods for carbon nanotubes (CNTs).[17-19] Moreover, the distinctive electronic properties and large aspect ratio of CNTs have made their combination with elastomer ideal for diverse applications including sensing, EMI shielding and flexible electronics,[20-23] which suggests that understanding their interfaces can be beneficial for material design towards target applications. The surfactant and coupling agent were used to construct different interfaces in CNTs/silicone elastomer nanocomposites. The effects of interfacial properties on dielectric response have been studied by dielectric spectroscopy. While dielectric measurements below the frequency of dozens or hundreds of megahertz have long been used for developing high-$k$ materials towards energy storage applications,[24, 25] rare attention has been given to the utilization of high frequency measurements as a possible tooling for exploring their correlation with structural features. Thereby, for the first time a relatively high frequency range in the order of gigahertz is employed in our study to interpret the meso-structural characteristics.

## 2. Results and Discussions

Three types of interfaces were constructed and CNTs/Silicone elastomer nanocomposites with various filler contents (0.12 vol%-0.96 vol%) were prepared by solution mixing method. As-received MWCNTs were first pre-treated with surfactant (Triton X-100) and coupling agent (KH550) before incorporated into polymer matrix, where Triton X-100 is a non-ionic molecule functions by physical absorption through π-π interaction and KH550 induces covalent bonding between CNTs and polymer matrix.[26, 27] In this way, interfacial properties can be tuned through physical interaction or chemical bonding. Silicone elastomer based nanocomposites contained raw MWCNTs, surfactant-treated MWCNTs and coupling agent-functionalized MWCNTs are denoted as MWCNTs/SE, AS-MWCNTs/SE, and CA-MWCNTs/SE respectively for convenience. SEM images of CNTs/Silicone elastomer samples (0.93-.94 vol%) containing pristine or modified CNTs are shown in the insets of **Figure 1**. Dotted and extracted CNTs were observed from the cutting surfaces of nanocomposites. In Figure 1(f) and (g) entangled CNTs are found to form aggregation in the nanocomposites, while in Figure 1(h) a relatively uniform dispersion of CNTs is displayed.

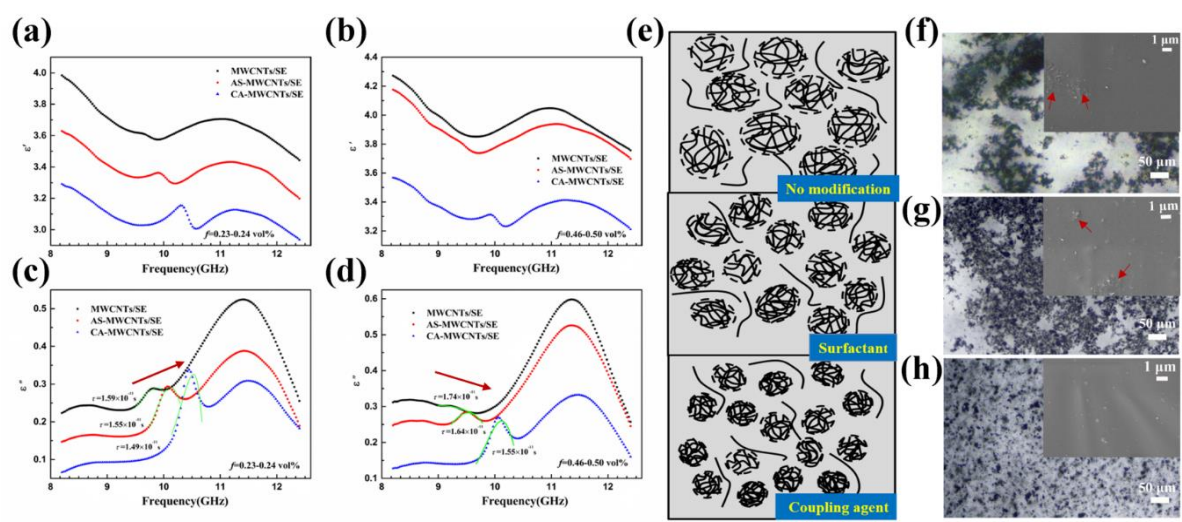

**Figure 1**. Frequency dependence of dielectric permittivity with different filler contents: (a) real part, $f$=0.23-.024 vol%. (b) real part, $f$=0.46-0.50 vol%. (c) imaginary part, $f$=0.23-.24 vol%. (d) imaginary part, $f$=0.46-0.50 vol%. Solid lines in (c) and (d) are fitting results for relaxation peak in the frequency range of 9-11 GHz using H-N function. (e) Schematic description of CNTs dispersion in silicone elastomer. (f)-(g) Optical microscopy and SEM images (insets) of CNTs dispersion in matrix before curing: (f) MWCNTs/SE. (g) S-MWCNTs/SE (h) CA-MWCNTs/SE.

Frequency dispersion is an intrinsic feature of dielectric spectra regardless of the various polarization mechanisms in dielectric materials,[28] which indicates that energy loss caused by a polarization process will be reflected in DS as frequency dependency of dielectric parameters. This can be mathematically interpreted by Kramers-Kronig relations in connecting the real and imaginary part of dielectric variables. To explore how dielectric spectroscopy can be used for understanding structural details in complex material system, we started off by examining how the dielectric dispersion of these nanocomposites varied with different interface modification methods. Figure 1(a)-(d) shows the complex dielectric frequency spectra of CNTs/Silicone elastomer with two representative filler contents (0.23 vol% and 0.46 vol%) in the frequency range of 8.2-12.4 GHz, in which a distinct relaxation peak was observed in the frequency range of 9-11 GHz for all samples. Dielectric spectra of the other two filler contents (0.12 vol% and 0.93 vol%) are displayed in the supporting information (Figure S1, Supporting Information), in which the variation of dielectric relaxation peak is insignificant. For nanocomposites with intermediate filler loadings of 0.23% and 0.46%, there is a blue shift of peak position after modification with surfactant and coupling agent as shown in Fig 1(b) and (d) by the arrow. This is the consequence of better dispersion and smaller agglomeration size of CNTs, corresponding to a smaller electric dipole size and meanwhile increased interfacial area, thereby both dipolar polarization and interfacial polarization are enhanced after modification. Triton X-100 has been reported to be capable of improving the dispersion of carbon nanotube by adsorption onto CNTs surfaces through π-π interaction and steric hindrance.[18, 29] Coupling agent is also shown to be effective for CNTs dispersion by surface functionalization and improve compatibility with both solvents and polymer matrix.[17] Figure 1(e) is a schematic diagram of dispersion and agglomeration state under different experimental conditions, where the CNTs/silicone elastomer modified by coupling agent achieves the best dispersion. This model is supported by the optical microscopy images (Figure 1(f)-(h)) presenting the dispersion scenario in nanocomposites.[19] Despite the clear agglomeration in Fig 1(f) and (g), it is shown in the optical image that CNTs modified by surfactant is more spatially extensive as compared to the unmodified ones, which is a sign of looser entanglement of CNTs fillers.[30] Unlike the previous two samples, CNTs were observed to be well-dispersed and dot-like in Fig 2(h), proving an even better dispersion as illustrated before.[18]

Such frequency dispersion is further analyzed with Havriliak–Negami relaxation model, which is an empirical function of non-Debye type dispersion.[31] The dielectric function takes the form:

$$\varepsilon^*(\omega)=\varepsilon_\infty+\frac{\Delta\varepsilon}{(1+(i\omega\tau)^\alpha)^\beta} \tag{1}$$

where $\varepsilon_\infty$ is the permittivity at high frequency limit, $\Delta\varepsilon=\varepsilon_s-\varepsilon_\infty$ and $\varepsilon_s$ is the static permittivity at low frequency. $\alpha$ and $\beta$ describes the asymmetric and symmetric distribution of dielectric spectra respectively, and $\tau$ is the characteristic relaxation time. With relaxation peak in the frequency range of 9-11 GHz in imaginary dielectric spectra fitted as shown in Figure 1(c) and (d) (solid curves), the corresponding dielectric relaxation times were calculated for each sample. For these two filler contents, the dielectric relaxation time decreases after modification by surfactant and coupling agent, which is the result of enhanced dielectric polarization contributed by smaller dipole size and enhanced interfacial polarization. On comparing the dielectric relaxation time for the samples with different filler contents prepared via same modification method, the relaxation time increases for all the three situations, which is caused by the relatively large dipole size due to worse dispersion at higher filler content. In other words, dipolar polarization dominates when the dispersion of nanoparticles plays a key role in determining the whole material system. Notably, the discrepancy of the relaxation time between samples with low and high filler contents become smaller after modification by both surfactant and coupling agent, suggesting that the increase of relaxation time caused by larger dipole size is partially weakened by the more important influence of interface to dielectric relaxation as the dispersion of nano-fillers is improved. Accordingly, the dielectric dispersion pattern of nanocomposites with low and high filler loading can be explained exactly in terms of changes in relative contribution by interfacial and dipolar polarization. When the dispersion of CNTs is very poor for the highest filler content (0.93-0.94 vol%) in this study, the accumulation of CNTs becomes so serious that the relaxation peak can even not be observed in the measured frequency range for MWCNTs/SE and AS-MWCNTs/SE nanocomposites (Figure S1(d), Supporting Information). For the lowest filler content where the dispersion is overall good and dipole size should be roughly similar for all the three methods, relaxation peak was not observed for MWCNTs/SE due to weak interface and filler-matrix interaction (Figure S1(c), Supporting Information), which further confirms that the relaxation peak is actually correlated to interfacial properties and interfacial polarization should be dominant as dispersion is not a concern.

It is also found in Figure 1(c) and (d) that the relaxation peak becomes clearer for AS-MWCNTs/SE and even sharper for CA-MWCNTs/SE nanocomposites, which is ascribed to better filler-matrix interaction and stronger interface. Dynamic mechanical analysis (DMA) was conducted for characterizing filler-matrix interaction (Figure S2, Supporting Information), and improved $T_g$ together with overlapped peaks in the tan $\delta$ curves after modification also demonstrates better dispersion and enhanced filler-matrix adhesion, [29, 32] which indicates that the relaxation peak in the frequency range of 9-11GHz in our experiments is indeed closely related to the interfacial interaction between nano-filler and the polymer matrix. Raman spectroscopy has also been used for interfacial analysis, where position change of characteristic peaks are shown to be induced by the different tangential stretching modes of the $sp^3$-hybridized and $sp^2$-hybridized carbons on CNTs surfaces influenced by interfacial molecules. [33] However, no obvious change was observed in Raman spectra of the three kinds of nanocomposites (Figure S2, Supporting Information), which is attributed to the fact that the interfaces constructed here by the limited active sites on pristine CNTs are relatively weak, [29] and hence cannot be effectively detected by Raman shift of the nanocomposites.

We notice that both real and imaginary part of complex permittivity are decreased after modification by surfactant or coupling agent for all the studied filler contents, which could be caused by the corresponding drop of electrical conductivity. [29, 32] The relative value change comparing to that of the unmodified samples were plotted in Figure 2(a) and (b). The reduction of permittivity caused by KH550 is more severe than that by Triton X-100. The relative change of permittivity with increasing filler content is quite distinct for surfactant and coupling agent treatment due to the different mechanisms of modification. The relatively strong bonding between CNTs and matrix (shown in Fig 2(c)) induced by KH550 could disturb the π-electron system of CNTs more seriously, thereby a remarkable decline with increasing filler content was observed. With increasing filler contents, the difference between the conductivity of the unmodified samples and that modified by KH550 becomes more pronounced, contributing to a relatively larger reduction of permittivity. Meanwhile, Triton X-100 can weaken charge transfer in CNTs by functioning as interfacial molecules through adsorption or wrapping the tube surface. Unlike CA-MWCNTs/SE nanocomposites, the fluctuation in the reduction degree for various filler contents could be caused by different modification extent from the dynamic adsorption and desorption process as illustrated in Fig 2(d). This dynamic process results from relatively weak interaction between Triton x-100 molecules and CNTs during the high speed centrifugal mixing process, leading to different relative change of conductivity.

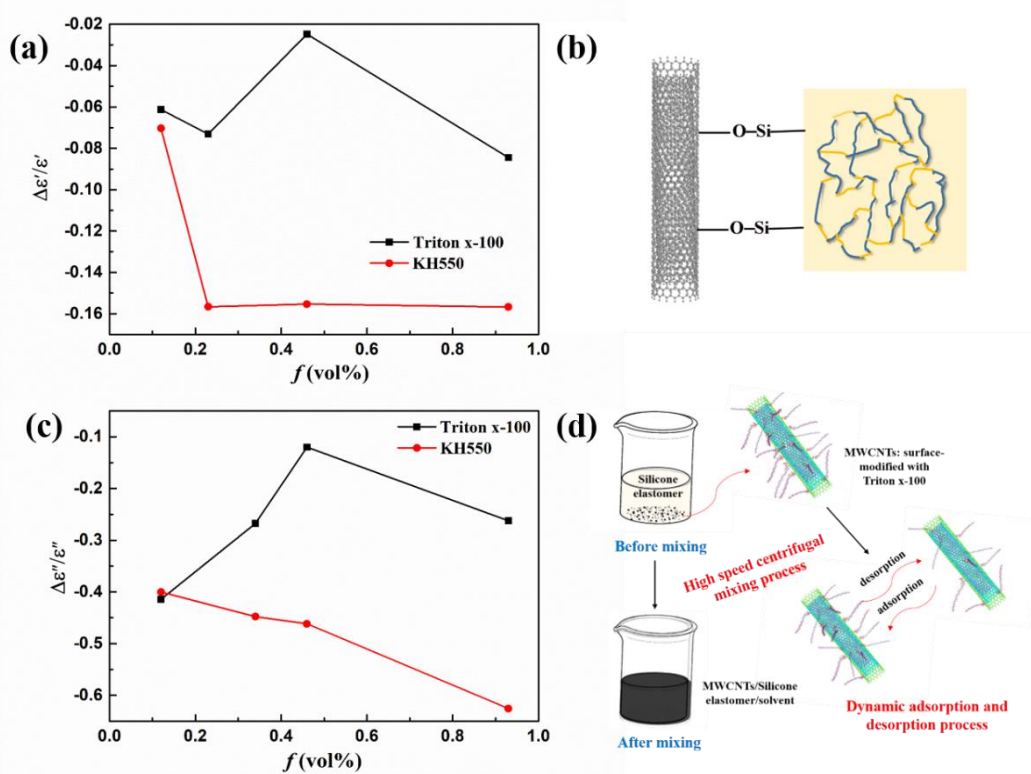

**Figure 2**. Relative variation of complex permittivity after interfacial modification: (a) real and (b) imaginary part. (c) Schematic description of coupling agent in bonding CNTs and polymer matrix. (d) Schematic illustration of desorption-adsorption process for samples modified by surfactant.

Based on previous analyses, engineering interface has a significant influence on the dielectric response of CNTs/silicone elastomer nanocomposite, where the structural difference concerning dispersion, agglomeration and properties of interfacial molecules have been reflected in terms of change in value, peak position and the amplitude of relaxation peaks. Looking back at the information presented above, it remains to be understood how interfaces functionalize and develop in nanocomposites. Considering the difficulty in characterizing the role of interface in functional nanocomposites directly, we introduced external forces to impact on interface during dielectric measurements and aim to formulate the real effect of interfaces in dielectric response. Additionally, it would be effective to characterize dynamic structural changes and the evolution of interface by introducing different cycles of loading.

To this end, we are inspired by the research on the mechanical properties and related structural changes of CNTs/polymer nanocomposites. It has been reported that mechanical loading would affect both the structural and electrical properties. Vertuccio et al. [34] explained that applied tensile strain could result in the widening of inter-particle distance and altering of conductive paths. Hydroxyl-functionalization of CNTs was found to increase the availability of nanotubes surface and thereby increase the stress transfer efficiency.[35] An increase of relative resistivity with increasing pressure was observed, and the breakage and formation of conductive network was found to be sensitive to the content of coupling agent at the interface[36]. Hence, integrating mechanical loading into dielectric measurements can be enlightening for investigating structural characteristics. As the interfaces constructed in this study is weak, a relatively small strain is determined in our study to emphasize the effect of interface and keep the effect of dispersion and distribution of CNTs minimized. Dielectric measurements were conducted immediately after load/unload cycling for all the studied cycles as plotted in Figure 3-5. It was found in the first place that peak positions of all three kinds of nanocomposites remained almost the same during cyclic loading, which suggests that the dispersion of CNTs maintain intact during this process.

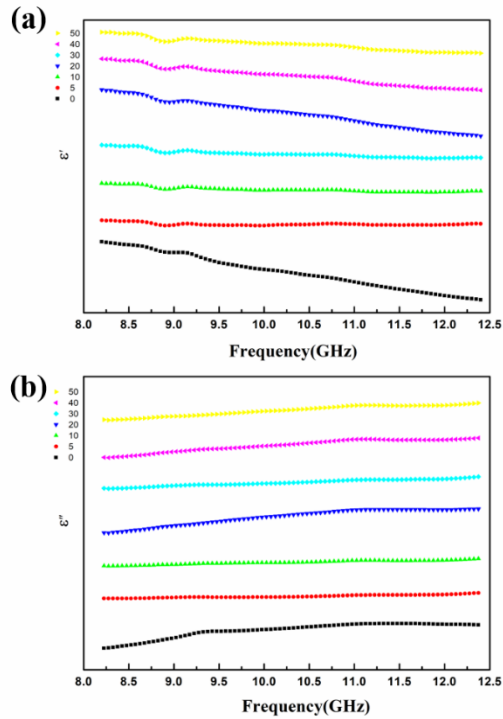

**Figure 3**. Variation of complex dielectric frequency spectroscopy (8.2 GHz-12.4 GHz) for MWCNTs/SE nanocomposites ($f$=0.50%) with cyclic tensile loading (5, 10, 20, 30, 40, 50 cycles, strain ranges from 26% to 31% for each cycle): (a) real part. (b) imaginary part.

Figure 3 shows the variation of dielectric frequency dispersion for MWCNTs/SE nanocomposites ($f$=0.46~0.50 vol%) with cyclic tensile loading, dielectric measurements were performed respectively after 5, 10, 20, 30, 40, 50 (accumulative number) cycles of tensile strain. The shape of dielectric spectra remains the same during the cyclic loading-dielectric measurements process. The characteristic relaxation peak appears at the frequency range of 9-9.5 GHz (highlighted in Figure 3 with the dotted straight line), which is consistent with the profiles in Figure 1(d). The relaxation peak disappears after 5 cycles of tensile loading and remained unchanged afterwards. As the relaxation peak correlated with interfacial properties for MWCNTs/SE is the weakest among the three kinds of samples, indicating that it is not sensitive to cyclic loading.

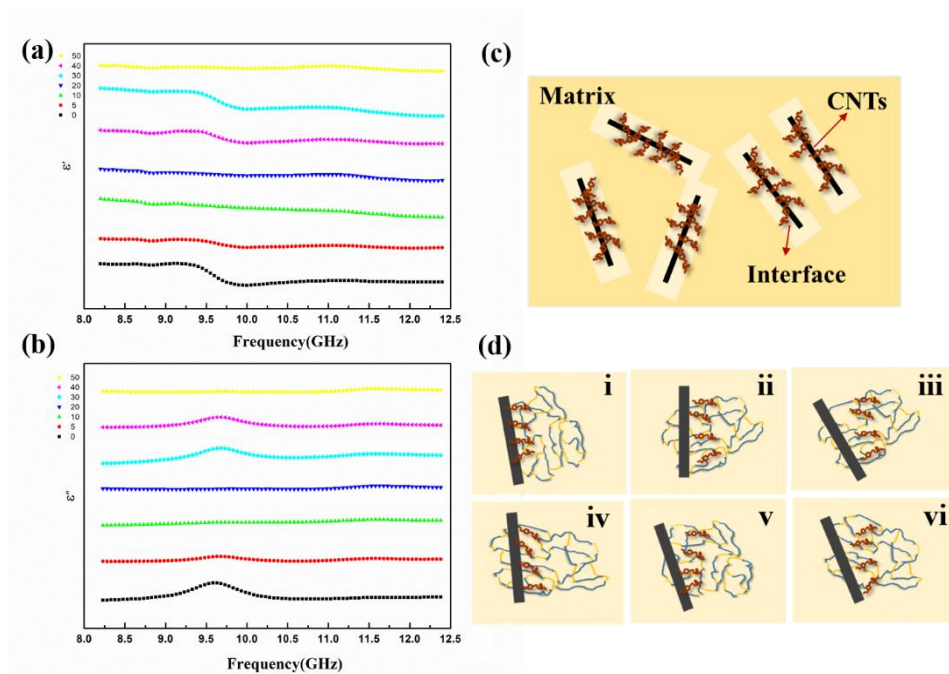

**Figure 4**. Variation of complex dielectric frequency spectroscopy (8.2 GHz-12.4 GHz) for S-MWCNTs/SE nanocomposites (f=0.46~0.50%) with cyclic tensile loading (5, 10, 20, 30, 40, 50 cycles, strain averages 28% for each set of cycles): (a) real part. (b) imaginary part. (c) Schematic illustration of AS-MWCNTs/SE nanocomposites with surfactant molecules on CNTs surface. (d) From i to vi: reconstruction process of surfactant molecules on CNTs surface with the increasing of cyclic loading times.

By contrast, some illuminating phenomena were observed for the other two types of composites with modified surfaces. For the sample with interface modified by Triton X-100, the relaxation peak remains in the frequency range of 9.5-10 GHz. The relaxation peak first weakens after 5 cycles of tensile strain and disappears after 10 cycles of loading. Interestingly, the peak reappears after 30 cycles of tensile strain, remains after 40 cycles of loading and vanishes after the last 10 cycles. The evolution of relaxation peak displays a reversible process with loading cycles, i.e., the destruction and reconstruction of surfactant-modified interface in nanocomposites (Figure 4(d)). In detail, the interface of AS-MWCNTs/SE has been formed by the entanglement of MWCNTs and polymer chains together with physical adsorption of Triton X-100 molecules in between as depicted in Figure 4(c). When the relatively small tensile strain was first applied to the materials, the slide between filler and matrix chains causes the desorption of the interfacial molecules from the CNTs surfaces (Figure 4(d)i-iii), resulting in the decrease of interface-related dielectric response. As the relatively small strain will not cause meaningful movement of CNTs, further cyclic loading on the material may lead to the re-attachment of the detached molecules on a different site on the CNTs surfaces (Figure 4(d)iii-iv). When the sample is submitted to tensile loading again, the re-detachment of molecules takes place (Figure 4(d)iv-vi), suppressing the relaxation peak after 50 cycles of loading. This structural change can also be interpreted from the perspective of energy. For such molecules like Triton X-100, the relatively large steric hindrance can dominate the overall stabilization,[37] making it important that the molecules should be in a proper site to lower energy. When the nanocomposite system is disturbed by applied loading, the adjustment of molecules attachment and its location would happen in order to maintain a relatively low energy state. In this way this destruction-reconstruction process is formed. It is of great interest that such effect of organic surfactant molecules could not only increase the compatibility of filler and matrix but explain the reversible interaction, yielding the corresponding changes illustrated above.

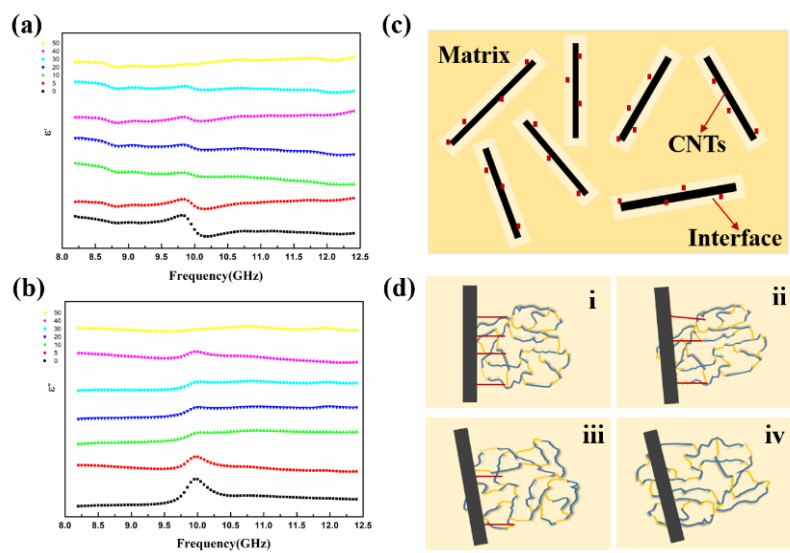

**Figure 5**. Variation of complex dielectric frequency spectroscopy (8.2 GHz-12.4 GHz) for CA-MWCNTs/SE nanocomposites (*f*=0.46) with cyclic tensile loading (5, 10, 20, 30, 40, 50 cycles, strain averages 28% for each set of cycles): (a) real part. (b) imaginary part. (c) Schematic illustration of CA-MWCNTs/SE nanocomposites with functionalities on CNTs surface, where dots on the surface of CNTs represents the interfacial functionalities formed by coupling agent and the light-colored parts around CNTs surfaces are the interfacial zone. (d) From i to iv: gradual de-bonding between CNTs and matrix chains with increasing number of loading cycles.

Unlike the former two samples, the CA-MWCNTs/SE nanocomposite presents the relaxation peak gradually decreased with strain cycles and finally vanished at 50 cycles (Figure 5(a) and (b)), which can be explained by the features of such interfaces with covalent bonding between filler and polymer chains. Such process of gradual change at the interface upon application of cyclic loading are modeled as Figure 5(d). The interface of as-prepared CA-MWCNTs/SE nanocomposites is depicted with several active sites enabled by coupling agent between CNT and polymer chains (Figure 4(d)i). With applied tensile strain, the relative position between CNT and polymer chains is changed, causing the loosening of the bonding in between (Figure 4(d)ii). This breakage of the covalent interaction becomes increasingly severe with loading times (Figure 4(d)ii-iii), and eventually the adhesion of the interface (Figure 4(d)iv) is ruined totally, leaving only the physical entanglement at the interface, together with the cancellation of interface-related relaxation peak in dielectric frequency spectroscopy. In other words, the gradual disappearance of relaxation peak is attributed to the accumulation effect of de-bonding and breakage at the interface.

## 3. Conclusion

In this study, investigation of dielectric frequency dispersion in the frequency range of 8.2-12.4 GHz has been highlighted for MWCNTs/silicone elastomer nanocomposites with different interfacial properties. The two most focused issues (interfacial interaction and dispersion) concerning different length scales for CNTs-reinforced nanocomposites are reflected and featured in the dielectric spectra in the form of variation in value, peak position and the amplitude of relaxation peaks. By analyzing dielectric relaxation mechanism from various perspectives, we have found that engineering interface is only effective when the challenge of dispersion is resolved for nanocomposites, which is enlightening for tuning properties elaborately and improving the performance of functional polymer nanocomposites. For example, since microwave absorption of CNTs composites is dependent on the dispersion and interfacial properties,[38] the dielectric response frequency range can be effectively tuned by adjusting the dispersion and interfacial interaction in nanocomposites with decorating molecules to modulate the operating frequency of microwave absorber. Hence delicate control of the interface and dispersion state based on the physical mechanism and fine chemical modification demands essential research efforts in order to improve composites performance.

By integrating mechanical measurements with high frequency dielectric testing, we have resolved the dynamic characteristic structural changes correlated with the interface and the role of interfacial properties (molecules, force types, degrees of interaction) in influencing dielectric relaxation phenomena. Key conclusions to draw from the results of dielectric-cyclic loading coupling are summarized as follows: (i) this method could be useful for characterizing relatively weak interfaces in CNTs nanocomposites and likely other nanocomposites with reasonable electrical or dielectric response; (ii) analyses of the destruction-reconstruction process in AS-MWCNTs/SE nanocomposites and the gradual breakage of interface in CA-MWCNTs/SE are instrumental to optimized design of interface in multifunctional nanocomposites. (iii) our study affords unique understanding of both static interfacial interactions and dynamic interface evolution induced by external stimuli of practical significance. As a final thought, if we introduce the concept of Material Genome Initiative, [39, 40] it is plausible to map the correlation of parameters describing structural features (dispersion state, interfacial interaction, filler properties) and characteristics of dielectric spectroscopy (value, peak position, amplitude of peaks) by applying high-throughput experimental and simulation methods at multiple length scales, thus making DS a more effective tool in smart selection and exploration of functional nanocomposites.

## 4. Experimental Section

*Materials*: Pristine MWCNTs (outer diameter: <8nm, inner diameter: 2-5nm, length:10-30um) grown by chemical vapor deposition (CVD) with a purity of 95% were purchased from Chengdu Organic Chemicals Co.,Ltd., Chinese Academy of Sciences. SYLGARD(R) 184 Silicone elastomer kit from Dow Corning Co. was used as polymer matrix. Triton X-100 (octylphenol ethylene oxide condensate, Laboratory Grade) with critical micelle concentration (CMC) from 0.22 to 0.24 mM was supplied by Sigma-Aldrich. 3-Aminopropyltriethoxysilane (KH550) were purchased from Adamas-beta. Pre-treatments of MWCNTs were first carried out by dispersing MWCNTs into Triton X-100/ ethanol (2.7 mg/ml) or KH550/ethanol solution (1 wt.%) and ultrasonicated for 30 minutes, followed by drying in oven at 50 ℃ .

*MWCNTs/silicone elastomer nanocomposites:* The MWCNTs/silicone elastomer nanocomposites were prepared by solution mixing method. Pristine and modified CNTs were first ultrasonicated in tytrohydrofuran (THF) solvent for 1h and meanwhile silicone elastomer was dissolved in THF, which were then mixed by a planetary centrifugal mixer. The THF solvent was removed by evaporation at 50℃ before the mixer was degassed and cured in the mold at 125℃ for 1h. The dimension of the mold is 22.86 mm×10.16 mm×5 mm. Several series of nanocomposite samples with different filler contents were prepared.

*Characterization：* Morphologies of MWCNTs/silicone elastomer were observed by field emission scanning electron microscopy (Zeiss, Utral 55). Dielectric measurements in the frequency range of 8.2-12.4 GHz were performed by vector network analyzer (R&S，ZNB20). Dispersion of CNTs were characterized using optical microscopy (Zeiss, Axio Vert.Al) after the solvent was evaporated. Raman spectroscopy was performed using a DXR Smart Raman spectrometer (irradiation wavelength: 532nm). Storage modulus and tan$\delta$ were determined using a dynamic mechanical analyzer (TAQ800). Tensile mode was used for testing from -130℃ to 80 ℃ at a heating rate of 5 ℃/min. Cyclic loading were carried out on a mechanical testing system (Instron 5943) at room temperature. Strain rate was set as 5mm/min and elongation for each cycle were determined as 3mm, the strain averages 28% (±3%) for each set of cycles.

Dielectric measurements were performed immediately after selected sets of cycles.

**Supporting Information**
Supporting Information is available from the Wiley Online Library or from the author.

**Acknowledgements**
FXQ would like to thank the financial support of NSFC No. 51501162 and No. 51671171, and National Youth Thousand Talent Program' of China.